\pdfoutput=1

\documentclass[11pt,twoside,a4paper,cmspaper,final,collab]{cms-tdr}
\def\svnVersion{46343:47304}\def\cmsCernNo{2001-007}\def\cmsCernDate{\today}\def\cmsMessage{Submitted to Physical Review Letters}

\begin{document}\cmsNoteHeader{SUS-10-002}

\hyphenation{had-ron-i-za-tion}
\hyphenation{cal-or-i-me-ter}
\hyphenation{de-vices}

\RCS$Revision: 47304 $
\RCS$HeadURL: svn+ssh://alverson@svn.cern.ch/reps/tdr2/papers/SUS-10-002/trunk/SUS-10-002.tex $
\RCS$Id: SUS-10-002.tex 47304 2011-03-24 17:49:18Z alverson $
\cmsNoteHeader{SUS-10-002} 
\title{Search for Supersymmetry in pp Collisions at $\sqrt {s}$ = 7 TeV in Events with Two Photons and Missing Transverse Energy}

\address[neu]{Northeastern University}
\address[fnal]{Fermilab}
\address[cern]{CERN}
\author[cern]{The CMS Collaboration}

\date{\today}

\abstract{
A search for supersymmetry in the context of general gauge-mediated (GGM) breaking  with the lightest neutralino
as the next-to-lightest supersymmetric particle and the gravitino as the lightest is presented. The data sample
corresponds
to an integrated luminosity of 36 pb$^{-1}$ recorded by the CMS experiment at the LHC.
The search is performed using events containing two or more isolated photons, at least one hadronic jet, and
significant missing transverse energy.
No excess of events at high missing transverse energy is observed.
Upper limits on the signal cross section for GGM supersymmetry between 0.3 and 1.1 pb at the 95\% confidence level are determined
for a range of squark, gluino, and neutralino masses, excluding supersymmetry parameter space that was
inaccessible to previous experiments.\\}

\hypersetup{%
pdfauthor={CMS Collaboration},%
pdftitle={Search for Supersymmetry in pp Collisions at sqrt(s) = 7 TeV in Events with Two Photons and Missing Transverse Energy},%
pdfsubject={CMS},%
pdfkeywords={CMS, supersymmetry}}

\maketitle

Supersymmetry (SUSY), in particular the version based on gauge-mediated breaking \cite{GGMa,GGMd2,GGMd3,GGMd4,GGMd5,GGMd1,GGMd},
is of particular theoretical interest for physics beyond the standard model (SM). Supersymmetry stabilizes the mass of the
SM Higgs boson, drives the grand unification of forces, and incorporates dark matter candidates within its framework.
Previous searches for SUSY with gauge-mediated breaking~\cite{GGMg,GGMh,lep1,lep2,lep3,lep4,h1} were performed using
a minimal model~\cite{snowmass}
as a benchmark. In that model, in order
to reduce the number of free parameters, several assumptions, including gaugino mass unification, are made. These assumptions lead to a
mass hierarchy in which
strongly interacting SUSY partners are much heavier than the lightest chargino and neutralino. For example, the current best lower limit on the neutralino
mass~\cite{GGMh} of 175 GeV corresponds to
gluino and squark mass limits of well above 1 TeV. In the more general case, the masses of strongly interacting SUSY partners can be much lighter, leading to large production
cross sections at the LHC, and allowing, even at low integrated luminosity,
for the exploration of parameter space inaccessible at previous colliders.

In this Letter we consider a general gauge-mediation (GGM) SUSY scenario~\cite{GGMe,GGMf}, with the gravitino as the
lightest  SUSY particle (LSP) and the lightest neutralino as the next-to-lightest (NLSP). In the following,
it is assumed that the neutralino decays promptly to a gravitino and a photon. Cases with either a large neutralino
lifetime or a large branching fraction into a \cPZ\ and a gravitino are not considered in this Letter.
The gravitino escapes detection, leading to missing transverse energy (\MET).  If R-parity \cite{rparity} is conserved,
strongly-interacting SUSY particles are pair-produced. Their decay chain includes one or several quarks/gluons,
and a neutralino, which in turn decays to a photon and a gravitino.
The topology of interest for this search is, therefore, two or more isolated photons with large transverse
energy (\ET), at least one hadronic jet, and large missing transverse energy \MET.

A detailed description of the CMS detector can be found elsewhere \cite{JINST}. The detector's central feature is a
superconducting solenoid providing a 3.8 T axial magnetic field along the beam direction.
Charged particle trajectories are measured by a silicon pixel and strip tracker system, covering
$0 \leq \phi \leq 2\pi$ in azimuth and $|\eta|<2.5$, where the pseudorapidity $\eta = - \ln{\tan{\theta/2}}$,
and $\theta$ is the polar angle with respect to the counterclockwise beam direction.
A lead-tungstate crystal electromagnetic calorimeter (ECAL) and a
brass/scintillator hadron calorimeter (HCAL) surround the tracker volume.
For the barrel calorimeter ($|\eta|<1.479$), the modules are arranged in projective towers.
Muons are measured in gas detectors embedded
in the steel return yoke of the magnet.
The detector is nearly hermetic, allowing for reliable measurement of \MET.
In the 2010 collision data, photons with energy greater than 20 GeV are measured within the barrel ECAL
with a resolution of better than 1\%, which is dominated by intercalibration precision.

The data used in this analysis, corresponding to an integrated luminosity
of 36 pb$^{-1}$, were recorded during the 2010 run at a center-of-mass energy ($\sqrt {s}$) of 7 TeV
at the Large Hadron Collider (LHC) at CERN.
A two-level trigger system was used, based on the
presence of at least one photon with a minimum transverse energy of 30 GeV.
This data sample is used for the selection of both signal candidates and control samples used for
background estimation. The efficiency for offline selected events to pass the trigger is estimated to be above 99\%.

The photon candidates are reconstructed from clusters of energy in the ECAL. Candidates are required to have
$\ET\geq 30$ GeV and $|\eta|\leq1.4$.
The ECAL cluster shape is required to be consistent with that expected from
a photon, and the energy detected in the HCAL behind the photon shower is required not to exceed 5\% of the ECAL energy.
To suppress photons originating from quark/gluon hadronization, the photons are required to be isolated from other activity
in the tracker, ECAL and HCAL.
The scalar sums of transverse energies of tracks and calorimeter deposits within
$\Delta R = \sqrt{ (\Delta\eta)^2+(\Delta\phi)^2} = 0.4$ of the candidate's direction are determined, after excluding
the contribution from the candidate itself. These isolation sums are
required to be $\leq 0.001 \times \ET + 2.0 $ GeV, $0.006 \times \ET + 4.2 $ GeV, and $0.0025 \times \ET + 2.2 $ GeV,
with \ET in GeV, for the tracker, ECAL, and HCAL, respectively.

Photons that fail either the shower shape or track isolation requirement are referred to as {\it fake photons}. Most of
these fake photons are quarks/gluons that have hadronized predominantly to $\pi^0$s decaying into photons.
They are used for the background estimation derived from data.

The criteria above are efficient for selection of both electrons and photons. To separate them reliably, we search for
hit patterns in the pixel detector consistent with a track from an electron (pixel match). The candidates with (without) pixel
match are considered to be electrons (photons) \cite{egam}. Events in the signal sample (referred to below as the $\gamma \gamma$ sample)
are required to have at least two photon candidates.

Jets are reconstructed from energy deposits in the calorimeters using the anti-$k_T$ clustering algorithm \cite{jets}
with a size parameter of 0.5.
The jet energy is corrected using reconstructed
tracks \cite{jet2}. Selected jets must have $\ET \geq$ 30 GeV and $|\eta| \leq 2.6$.
In addition, to separate real jets from anomalous HCAL signals, the fraction of energy
contributed to the jet shower by the highest energy HCAL channels must be $\leq$ 98\%, the jet must have no single  HCAL channel
containing more than 90\% of its total energy, and finally the ECAL energy fraction of the jet must be $\geq$~1\%. To be retained
in the signal sample,
events must contain at least one such jet isolated from both of the two highest-\ET photon candidates by $\Delta R \geq 0.9$.

The \MET is determined using calorimeter energy deposits. Corrections are applied by replacing calorimeter tower energies matched
to charged hadrons and muons with their corresponding charged-track momenta ~\cite{MET}.

Events are generated in the benchmark SUSY model~\cite{NLSP} using {\PYTHIA 6.4}~\cite{pythia} in a three-dimensional grid of the NLSP, gluino, and squark masses.
The soft masses of the squarks are taken to be degenerate. Sleptons and all other gauginos except the NLSP are assigned a mass of 1.5 TeV.
The QCD production cross section at NLO is calculated for these points using {\PROSPINO 2.1}~\cite{prospino}, and is dominated
by gluino-gluino, gluino-squark, and squark-squark production.
The generated events are passed through the CMS
detector simulation program, which is based on the {\GEANTfour}~\cite{geant} package, and reconstructed using the same
program as for the collision data to ensure that all
features of the data, such as trigger and reconstruction, are applied to the Monte Carlo (MC) simulated SUSY signal sample.

The SUSY signal can be mimicked in several ways. Irreducible backgrounds from SM processes such as
$ \cPZ(\rightarrow \nu \bar{\nu})\gamma\gamma$ and $\PW(\rightarrow \ell \nu) \gamma\gamma$ are negligible.
The main backgrounds
arise from SM processes with misidentified photons and/or mismeasured \MET.
The dominant contribution comes
from mismeasurement of \MET in QCD processes such as direct diphoton, photon plus jets, and multijet production, with jets
mimicking photons in the latter two cases. This background is referred to as the {\emph {QCD background}}. The strategy for
determining this background is to use control samples that are kinematically similar to the candidate sample and that can be
reasonably assumed to have
no significant genuine \MET. Two such samples are used. The first is a sample containing two fake photons, referred to as the
\emph {ff} sample
below, comprising QCD multijet events. The second (\emph {ee}) sample contains events with two electrons~\cite{electron}
with invariant mass between
70 and 110 GeV, and is dominated by $\cPZ\rightarrow \Pe\Pe$ decays.

The \ET resolution for electrons and fake photons is similar to the resolution for photons and is much better
than the resolution for hadronic energy, so the \MET resolution is dominated by the latter.  The events in both control samples
are reweighted to reproduce the diphoton transverse energy distribution in the candidate $\gamma \gamma$ sample and, therefore,
the transverse energy
of the hadronic recoil against the diphoton system; the reweighting factors range from 0.3 to 1.7. The shapes of the \MET distributions
obtained from both control samples are identical within the statistical and systematic uncertainties and are used to predict
the \MET background by
normalizing to the number of  events with \MET $< 20$ GeV in the candidate sample.

The second background comes from events with real \MET. It is dominated by events with a genuine or fake photon and a $\PW$ that
decays into a neutrino and an electron, with the latter misidentified as a photon. This background is referred to as the
{\emph {electroweak background}}. An electron is
misidentified as a photon if it satisfies all the photon selection criteria but has no matching hit pattern in the pixel detector.
To model this background, an $\Pe\gamma$ candidate sample is defined, selected the same way as the $\gamma\gamma$ sample but requiring at
least one electron and at least one photon instead of at least two photons.
This sample is thus enriched with $\PW\gamma$ and $\PW$ plus jets events
(with real \MET) similar to those in the candidate $\gamma \gamma$ sample.  The probability, $f_{\Pe\rightarrow\gamma}$, to
misidentify an electron as a photon is measured with $\cPZ\rightarrow \Pe\Pe$ events to be (1.4 $\pm$ 0.4)\%.
The $\Pe\gamma$ sample is then weighted by $f_{\Pe\rightarrow\gamma}/(1-f_{\Pe\rightarrow\gamma})$ in order
to estimate the contribution of this background to the signal sample.

Finally, high-energy muons from cosmic rays or beam halo can deposit a large amount of energy in the ECAL, leading to events with
two photon candidates and \MET. These events are suppressed to a negligible level by the jet requirement described above.

\begin{figure}[hbtp]
\begin{center}
\includegraphics[width=0.95\columnwidth]{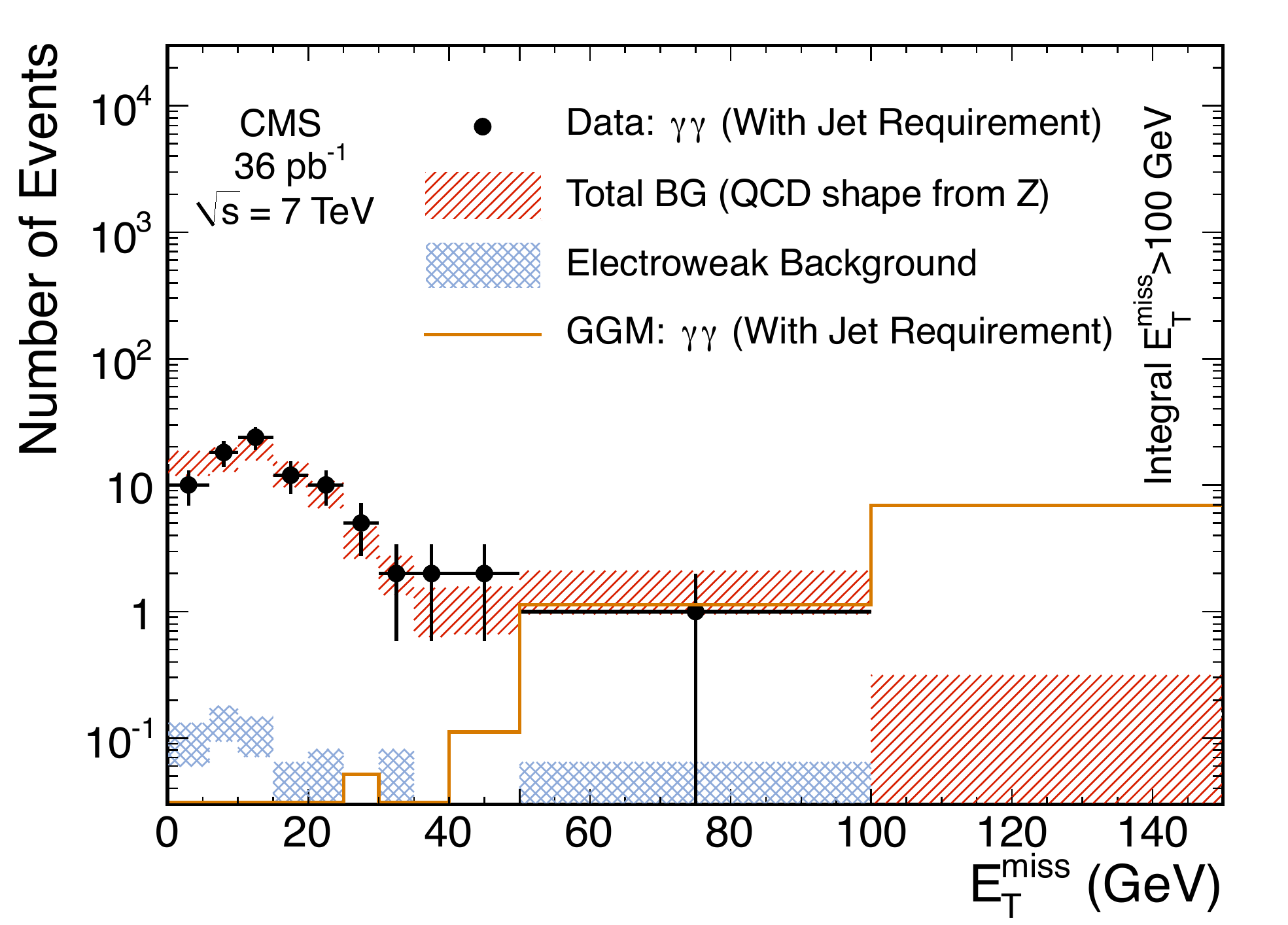}
\caption {\MET distribution for $\gamma\gamma$ data, including the jet requirement,  compared with backgrounds and a possible GGM SUSY signal.
The solid circles with error bars represent the data.  The double-hatched blue band represents the contribution of the electroweak background.
The single-hatched red band shows the sum of the electroweak background with the QCD \MET prediction obtained from the $\cPZ\rightarrow \Pe\Pe$ sample.
The widths of the bands correspond to the sum of the statistical and systematic uncertainties on the backgrounds.
The prediction of the GGM SUSY sample point described in the text is shown in the plot as the solid line histogram.}
\label{close}
\end{center}
\end{figure}

The \MET distribution in the $\gamma\gamma$ sample is shown in Fig.~\ref{close}, together with the estimates of the
electroweak background and the total
background using the QCD prediction from $\cPZ\rightarrow \Pe\Pe$ and their uncertainties.
Events with \MET$\leq$ 20 GeV have
negligible SUSY signal contribution (0.022 $\pm$ 0.009 event for the GGM SUSY sample point described below). The \MET distributions
for the \emph {ff} and \emph {ee} samples are then scaled so that their integrals
below 20 GeV match that of the $\gamma\gamma$ sample minus the estimated electroweak contribution.
Their integrals above 50 GeV give predictions of the QCD background.

Table~\ref{closurevt} summarizes the number of $\gamma\gamma$ events observed and the numbers of background events expected
with \MET$\geq$ 50 GeV. The statistical and systematic uncertainties on the background predictions due to reweighting and normalization are
shown separately. One event is observed, while the total
background is expected to be 0.53 $\pm$ 0.37 (1.71 $\pm$ 0.64) events using the \emph {ff} (\emph {ee})
samples and including the electroweak background. These two consistent estimates are averaged,
using log-normal distributions as probability density functions while taking into account the common component from the
electroweak background and the correlated uncertainty due to
the normalization, to obtain a prediction of 1.2 $\pm$ 0.4 background events. An additional conservative systematic
uncertainty of 0.7 events is assigned
by taking the largest difference between the average and the individual measurements, resulting in a total background
uncertainty of 0.8 events.

\begin{table*} [hbtp]
\caption{\label{closurevt} The number of events with \MET $\geq$ 50 GeV from the $\gamma\gamma$ event sample as well as the
predicted
number of background events with \MET$\geq$ 50 GeV using either the fake-fake events or the $\cPZ\rightarrow \Pe\Pe$ data.}
\begin{center}
\begin{tabular}{|l|c|c|c|c|}
\hline\hline
Type &  Number of & Stat & Reweight& Normalization\\
     &   events  & error& error   &   error \\
\hline
$\gamma\gamma$ events      &1 & & & \\
Electroweak background estimate   &$0.04\pm0.03$ &$\pm0.02$&$\pm0.0$&$\pm0.01$\\

QCD background estimate (\emph{ff})         &$0.49\pm0.37$ &$\pm0.36$&$\pm0.06$&$\pm0.07$\\

QCD background estimate (\emph {ee}) &$1.67\pm0.64$
&$\pm0.46$&$\pm0.38$&$\pm0.23$\\

Total background  (using \emph {ff})  &$0.53\pm0.37$ & & &\\
Total background  (using \emph {ee})  &$1.71\pm0.64$    & & &\\
Combined total background  &$1.2\pm0.8$    & & &\\
Expected from GGM sample point  &$8.0\pm1.7$    & & &\\
\hline\hline
\end{tabular}
\end{center}
\end{table*}

The efficiency for SUSY events to satisfy the selection criteria is determined by applying correction factors derived from the data
to the MC simulation of the signal.  Since there exists no large clean sample of isolated photons in the data, we rely on similarities
between the detector response to electrons and photons to extract the photon efficiency.
The difference between the efficiencies for electrons and photons according to the MC simulation is 0.5\%.
The ratio of the electron efficiency from $\cPZ\rightarrow \Pe\Pe$ events that pass all photon identification
criteria (except for the pixel match) to the corresponding electron MC efficiencies gives an
MC efficiency scale factor of 0.967 $\pm$ 0.015.
This scale factor is applied to the efficiency to identify a photon from NLSP decay obtained with the MC simulation.
The error on the scale factor also includes
possible systematic effects from multiple interactions per bunch crossing (pile-up), estimated to be less than 1\%.
The additional uncertainty on the efficiency of the requirement for no pixel match for a photon is estimated by varying the
amount of tracker material in the detector simulation and is found to be equal to 0.5\%. Other sources of systematic uncertainties
in signal yield include the uncertainty on
the integrated luminosity (11\%) \cite{lumi}, parton distribution function (PDF) uncertainty (10-40\%), and renormalization scale uncertainty
(10-20\%), depending on the SUSY masses.

\begin{figure}[hbtp]
\begin{center}
\includegraphics[width=0.95\columnwidth]{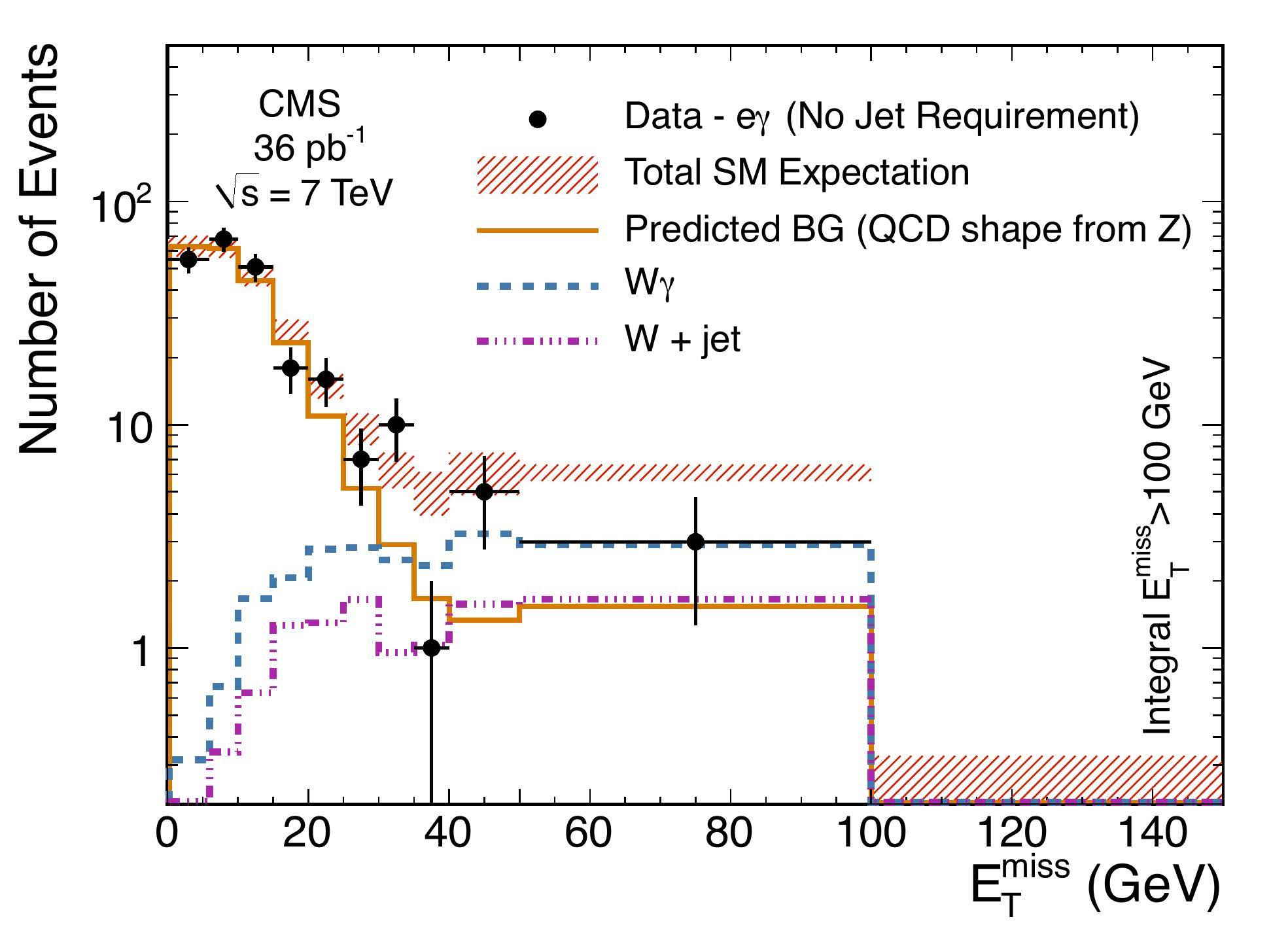}
\caption{\label{W} The \MET distribution of the $\Pe\gamma$ candidates (solid circles with error bars) is compared to the QCD expectation for this
spectrum from $\cPZ\rightarrow \Pe\Pe$ events (solid histogram).
The $\PW\gamma$ and $\PW$ plus jet (with the jet misidentified as a photon) contributions are shown as dashed and dash-dotted histograms,
 respectively.
The hatched red band represents the sum of all expected background components, and its width corresponds to the sum of the statistical and systematic
uncertainties on the backgrounds.}
\end{center}
\end{figure}

As a cross check of the analysis method, in particular the determination of the QCD background contribution,
the procedure is applied to a data sample selected in the same way as the $\Pe\gamma$ sample described above, except that there is
no requirement on the number of jets.
The check consists of estimating the QCD background, events with no true \MET, in a sample populated with both QCD
and events from the two dominant SM processes, $\PW\gamma$ and $\PW$ plus~jets, which have true \MET.
An excess in the observed number of events with large \MET over the estimated QCD background should be consistent with the yield
from the $\PW$ processes.
The \MET spectrum of the $\Pe\gamma$ events is shown in Fig.~\ref{W} and exhibits a clear deviation from
the predicted QCD background alone but agrees well with the sum of the QCD and $\PW$ backgrounds.

\begin{figure}[hbtp]
\begin{center}
\includegraphics[width=0.95\columnwidth]{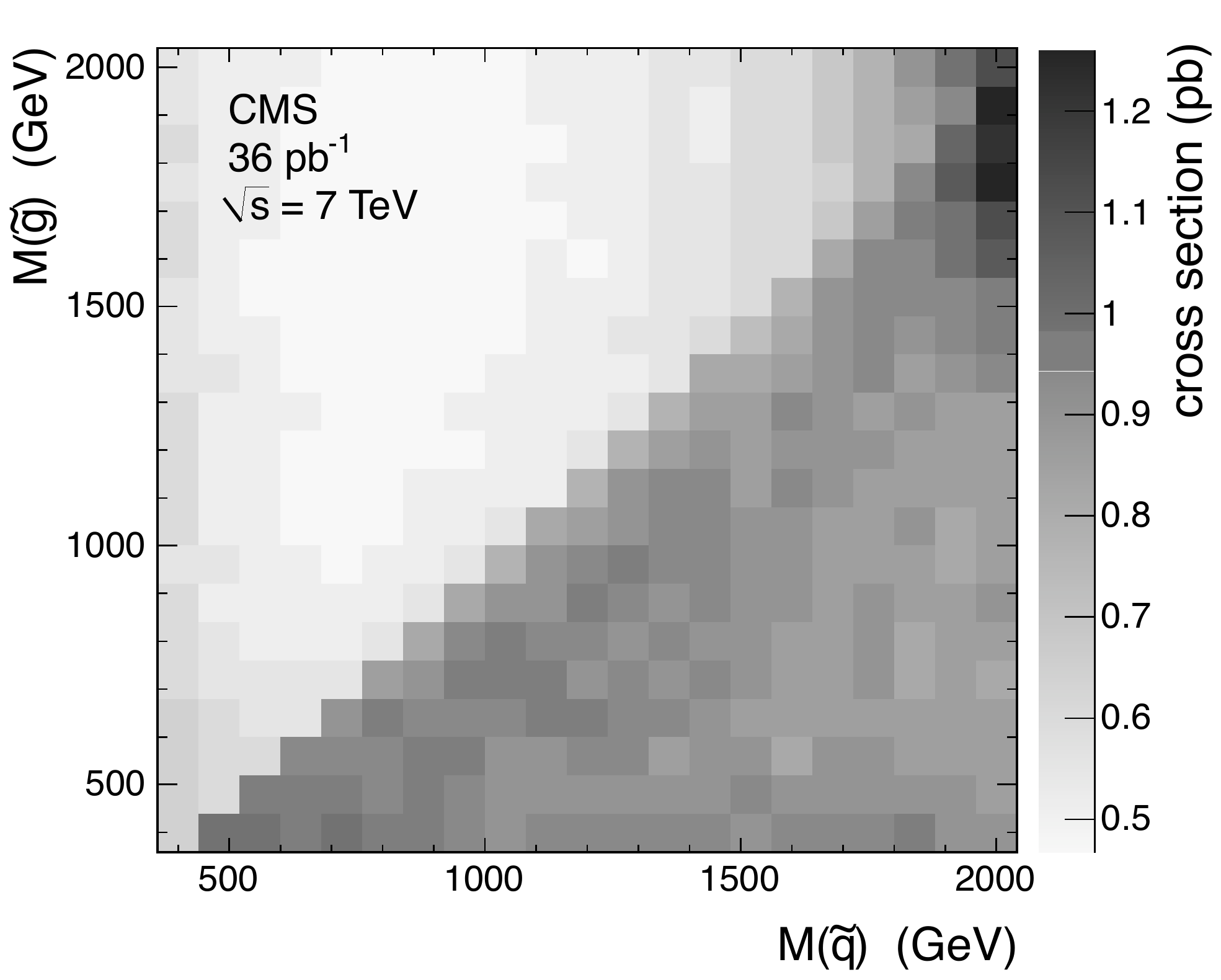}
\caption{\label{GGM-contours} 95\% CL upper limits for GGM production cross section as a function of squark ($\tilde q$) and
gluino ($\tilde g$) masses for a neutralino mass of 150 GeV.}
\end{center}
\end{figure}

This measurement and the estimated acceptance times efficiency are used to set upper limits on the gluino and squark
production cross sections, employing a Bayesian method described in Ref.~\cite{bay}. Both the log-normal and
gamma priors are used to integrate over nuisance parameters in order to incorporate uncertainties on the total
background rate, integrated luminosity, and total acceptance times efficiency.
The observed 95\% confidence level (CL) cross section limits, shown in Fig.~\ref{GGM-contours},
vary between 0.3 and 1.1 pb for a neutralino mass of 150 GeV as a function of squark and gluino masses. The variation
in the limits is due to the dependence of the photon isolation efficiency on the number of jets in the event.

To illustrate the limit-setting procedure, a particular signal point, with a squark mass of 720~GeV, gluino
mass of 720~GeV, and neutralino mass of 150~GeV, is considered. The NLO signal cross section for this point is
1.04 pb, the efficiency times acceptance is 0.203 $\pm$ 0.004 (stat) $\pm$ 0.008 (syst), and the PDF and scale uncertainties
are each 13\%.
The upper limit obtained from this measurement is 0.585~pb, while the expected cross section upper limit
at the 95\% CL is 0.628 pb. The expected number of events for this
SUSY sample point for an integrated luminosity of 36 pb$^{-1}$ is 8.0 $\pm$ 1.7 events. For comparison, the total NLO SUSY
production cross section at the Tevatron for the same model point is 0.3 fb, leading to an expectation of less than half of
an event for analysis~\cite{GGMh}.

\begin{figure}[hbtp]
\begin{center}
\includegraphics[width=0.95\columnwidth]{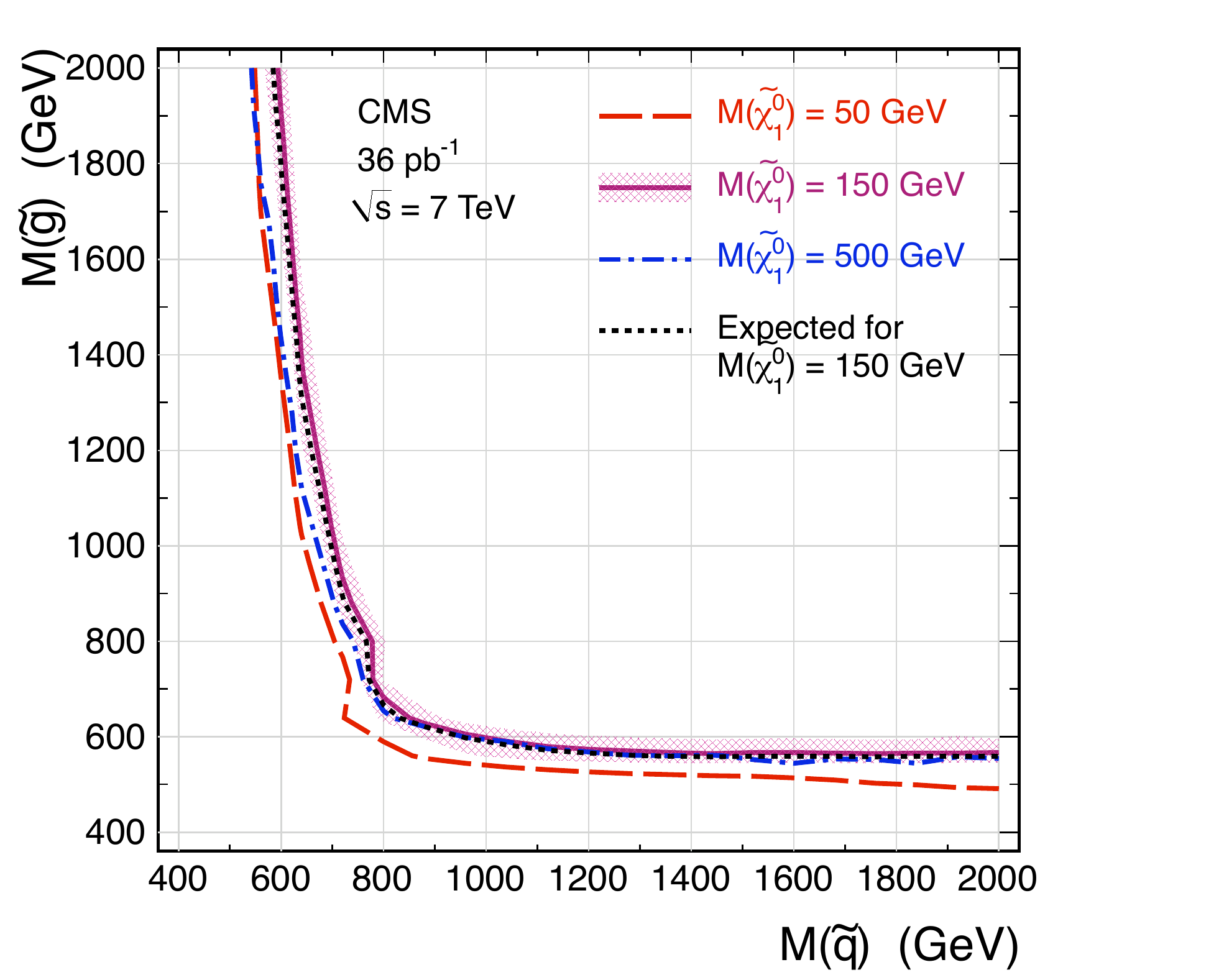}
\caption{\label{exclusion} Lower 95\% CL exclusion limits on the squark ($\tilde q$) and gluino ($\tilde g$) masses in
the GGM benchmark model for
50, 150, and 500 GeV neutralino ($\tilde\chi^0_1$) masses. The areas below and to the left of the lines are excluded.
The expected exclusion limit for 150 GeV neutralino mass is shown by the dashed line. The shaded band
represents {$\pm 1$} standard deviation of theoretical uncertainty on the GGM cross section.}
\end{center}
\end{figure}

The benchmark GGM model used in this Letter can be used to interpret these cross section
limits as lower limits on squark and gluino masses.
For each mass point we compare the predicted cross section with the measured upper limit and claim exclusion if the former
is greater than the latter. The exclusion contours for three different choices
of neutralino mass  are shown in Fig.~\ref{exclusion}, together with the expected exclusion limit for a 150 GeV neutralino mass.
The predicted cross section has uncertainties due to the choice of PDF and scale. We varied the predicted cross section by one standard
deviation of its uncertainty to ascertain the impact on the exclusion region. The results for 150 GeV neutralino mass are presented
as a shaded band around the central value of the exclusion contour.

In summary, a search for evidence of GGM SUSY production in events that contain two or more high transverse
energy isolated photons, one or more jets, and large \MET is presented.  No such evidence is observed, leading to upper limits
on the GGM SUSY cross section between 0.3 and 1.1 pb
at the 95\% CL across the parameter space of a benchmark model, setting the world's best direct
lower limits on squark and gluino masses.

We wish to congratulate our colleagues in the CERN accelerator
departments for the excellent performance of the LHC machine. We thank
the technical and administrative staff at CERN and other CMS
institutes, and acknowledge support from: FMSR (Austria); FNRS and FWO
(Belgium); CNPq, CAPES, FAPERJ, and FAPESP (Brazil); MES (Bulgaria);
CERN; CAS, MoST, and NSFC (China); COLCIENCIAS (Colombia); MSES
(Croatia); RPF (Cyprus); Academy of Sciences and NICPB (Estonia);
Academy of Finland, ME, and HIP (Finland); CEA and CNRS/IN2P3
(France); BMBF, DFG, and HGF (Germany); GSRT (Greece); OTKA and NKTH
(Hungary); DAE and DST (India); IPM (Iran); SFI (Ireland); INFN
(Italy); NRF and WCU (Korea); LAS (Lithuania); CINVESTAV, CONACYT,
SEP, and UASLP-FAI (Mexico); PAEC (Pakistan); SCSR (Poland); FCT
(Portugal); JINR (Armenia, Belarus, Georgia, Ukraine, Uzbekistan); MST
and MAE (Russia); MSTD (Serbia); MICINN and CPAN (Spain); Swiss
Funding Agencies (Switzerland); NSC (Taipei); TUBITAK and TAEK
(Turkey); STFC (United Kingdom); DOE and NSF (USA).

\bibliography{auto_generated}

\cleardoublepage\appendix\section{The CMS Collaboration \label{app:collab}}\begin{sloppypar}\hyphenpenalty=5000\widowpenalty=500\clubpenalty=5000\textbf{Yerevan Physics Institute,  Yerevan,  Armenia}\\*[0pt]
S.~Chatrchyan, V.~Khachatryan, A.M.~Sirunyan, A.~Tumasyan
\vskip\cmsinstskip
\textbf{Institut f\"{u}r Hochenergiephysik der OeAW,  Wien,  Austria}\\*[0pt]
W.~Adam, T.~Bergauer, M.~Dragicevic, J.~Er\"{o}, C.~Fabjan, M.~Friedl, R.~Fr\"{u}hwirth, V.M.~Ghete, J.~Hammer\cmsAuthorMark{1}, S.~H\"{a}nsel, M.~Hoch, N.~H\"{o}rmann, J.~Hrubec, M.~Jeitler, G.~Kasieczka, W.~Kiesenhofer, M.~Krammer, D.~Liko, I.~Mikulec, M.~Pernicka, H.~Rohringer, R.~Sch\"{o}fbeck, J.~Strauss, F.~Teischinger, P.~Wagner, W.~Waltenberger, G.~Walzel, E.~Widl, C.-E.~Wulz
% ...
\vskip\cmsinstskip
% ...
1:~~Also at CERN, European Organization for Nuclear Research, Geneva, Switzerland\\
% ...

\end{sloppypar}
\end{document}